\documentclass[journal=apchd5,manuscript=letter,layout=twocolumn]{achemso}

\usepackage[T1]{fontenc} 
\usepackage[exponent-product = \cdot, separate-uncertainty = true]{siunitx}
\usepackage{color} 

\author{Christopher Damgaard-Carstensen}
\email{cdc@mci.sdu.dk}
\author{Martin Thomaschewski}
\author{Fei Ding}
\author{Sergey I. Bozhevolnyi}
\email{seib@mci.sdu.dk}
\affiliation[SDU]{Centre for Nano Optics, University of Southern Denmark, Campusvej 55, DK-5230 Odense M, Denmark}

\title{Electrical Tuning of Fresnel Lens in Reflection}

\keywords{Metasurface, flat optics, electrical tunability, lithium niobate, Fresnel lens}

\begin{document}


\begin{abstract}
Optical metasurfaces have been extensively investigated, demonstrating diverse and multiple functionalities with complete control over the transmitted and reflected fields. Most optical metasurfaces are however static, with only a few configurations offering (rather limited) electrical control, thereby jeopardizing their application prospects in emerging flat optics technologies. Here, we suggest an approach to realize electrically tunable optical metasurfaces, demonstrating dynamic Fresnel lens focusing. The active Fresnel lens (AFL) exploits the electro-optic Pockels effect in a 300-nm-thick lithium niobate layer sandwiched between a continuous thick and nanostructured gold film serving as electrodes. We fabricate and characterize the AFL, focusing 800-\SI{900}{\nano\meter} radiation at the distance of \SI{40}{\micro\meter} with the focusing efficiency of \SI{15}{\percent} and demonstrating the modulation depth of \SI{1.5}{\percent} with the driving voltage of \SI{\pm10}{\volt} within the bandwidth of $\sim\!\SI{4}{\mega\hertz}$. We believe that the electro-optic metasurface concept introduced is useful for designing dynamic flat optics components. 
\end{abstract}


\section{Introduction}
Over the last decade, optical metasurfaces, representing nm-thin planar arrays of resonant subwavelength elements, have been extensively investigated, demonstrating diverse and multiple functionalities that make use of the available complete control over the transmitted and reflected fields \cite{Yu2011,Hsiao2017,Chen2016,Ding2017}. This progress led to the realization of numerous flat optical components in concert with the current trend of miniaturization in photonics. Large flexibility in the design of optical metasurfaces enabled numerous demonstrations of various functionalities, including beam-steering \cite{Pors2013_BS,Li2015_BS,DamgaardCarstensen2020_BS}, optical holograms \cite{Chen2013_OH,Wen2015_OH,Huang2015_OH}, and planar lenses \cite{Ding2019_PL,Yi2017_PL,Boroviks2017_PL}. Most of the developed optical metasurfaces are however static, featuring well-defined optical responses determined by the configuration of material and geometrical parameters that are chosen by design and set in the process of fabrication. Realization of dynamic metasurfaces faces formidable challenges associated with the circumstance that metasurfaces are fundamentally very thin, i.e., of subwavelength thickness, limiting thereby severely the interaction length available. Efficient tunability can be achieved through material property (phase) transitions or structural reconfigurations that result in very large refractive index changes, but these effects are inherently slow \cite{Che2020,vandeGroep2020,Park2020,Shirmanesh2020}. The speed limitations jeopardize the application prospects in emerging technologies, such as light detection and ranging (LIDAR) and computational imaging and sensing \cite{Schwarz2010,Jung2018}. 

The electro-optic Pockels effect enables fast electrically controlled modulation of material properties in several active media, e.g. lithium niobate (LN), electro-optic polymers or aluminum nitride \cite{Thomaschewski2020,Zhang2018,Smolyaninov2019}. Especially LN offers an attractive platform, due to its large electro-optic coefficients ($r_{33} = \SI{31.45}{\pico\meter/\volt})$, superb chemical and mechanical stability resulting in long-term reliability, and wide optical transparency range (0.35 - \SI{4.5}{\micro\meter}) \cite{Weis1985}. The aforementioned limitations in the available interaction length makes however exploiting comparatively weak electro-optic material effects problematic, resulting in rather weak tunability and modulation efficiency \cite{Zhang2018,Gao2021}. 

We introduce in this work an approach to realize electrically tunable optical metasurfaces by utilizing the electro-optic effect in a thin LN layer sandwiched between a continuous thick bottom and nanostructured top gold film serving as electrodes. Our approach is based on electrically tuning the light reflectivity near a high-fidelity Fabry-Perot resonance. This concept is implemented in dynamic (electrically controlled) Fresnel lens focusing. By conducting detailed numerical simulations and experiments for a 300-nm-thick LN layer, we demonstrate that the active Fresnel lens (AFL) exhibits tunable focusing and modulation in reflection at near-infrared wavelengths. The fabricated AFL is found to exhibit focusing of 800-\SI{900}{\nano\meter} radiation at the distance of \SI{40}{\micro\meter} with the focusing efficiency of \SI{15}{\percent} and modulation depth of \SI{1.5}{\percent} (for the driving voltage of \SI{\pm10}{\volt}) within the bandwidth of $\sim\!\SI{4}{\mega\hertz}$. We believe that the introduced electro-optic metasurface concept is useful for designing dynamic flat optics components.

\section{Results and Discussion}

\begin{figure}[!tb]
	\centering
	\includegraphics{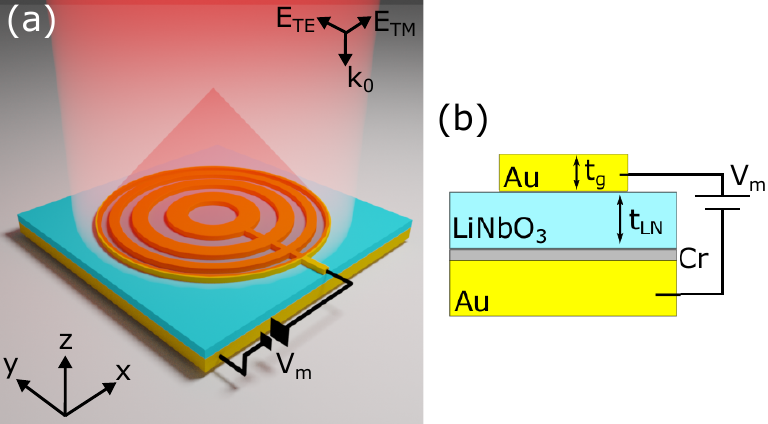}
	\caption{Schematics of the designed active Fresnel lens (AFL). (a) Three dimensional rendering of the zone plate showing focusing of the incident light under an applied voltage. (b) Cross section sketch displaying a semi-transparent gold ring deposited on a lithium niobate thin film, adhered to a gold back-reflector by a thin chromium adhesive layer.}
	\label{fig1:Schematic}
\end{figure}

\begin{figure*}[!tb]
	\centering
	\includegraphics{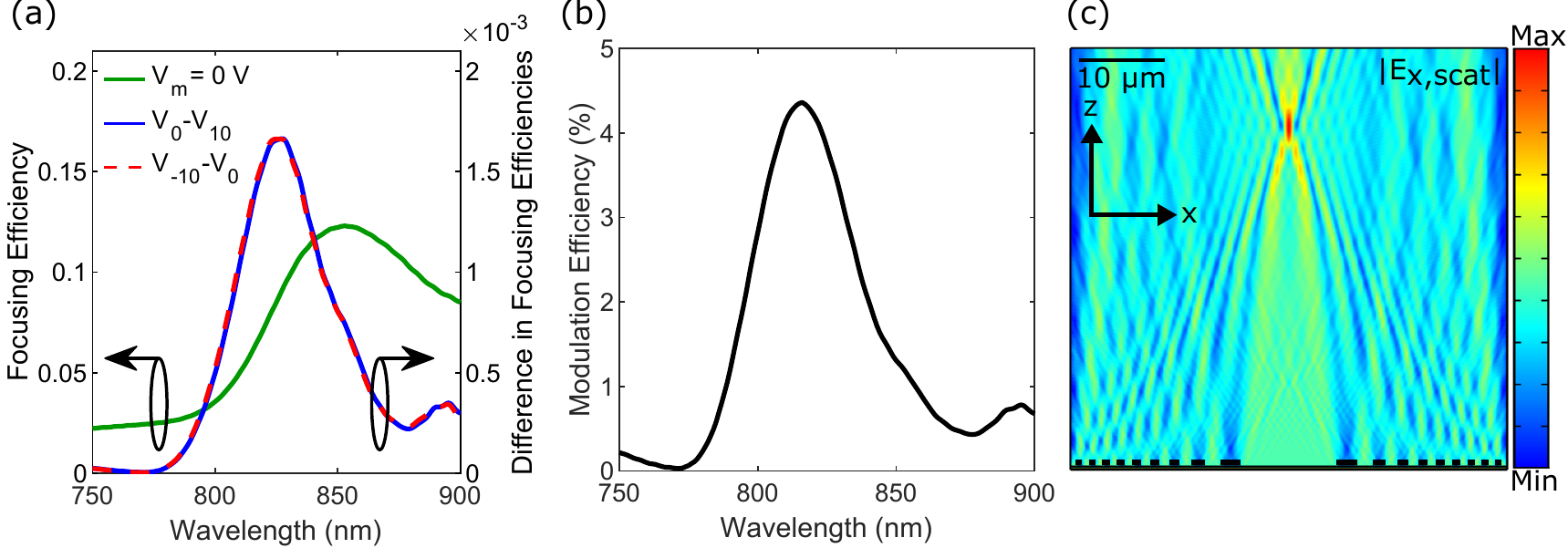}
	\caption{Calculated performance of the AFL. (a) Calculated focusing efficiency (left axis) as a function of wavelength without an applied modulation voltage, and variation in focusing efficiency (right axis) when applying a DC modulation voltage of \SI{\pm10}{\volt}. (b) Calculated modulation of the focusing efficiency as a function of wavelength, when applying a DC modulation voltage of \SI{\pm10}{\volt} (c) Calculated \textit{x}-component of the scattered field above an AFL with a designed focal length of \SI{40}{\micro\meter} at an incident wavelength of \SI{815}{\nano\meter} for a DC modulation voltage of \SI{-10}{\volt}. }
	\label{fig2:Simulations}
\end{figure*}

Figure \ref{fig1:Schematic} shows schematics of the proposed structure consisting of semi-transparent gold rings deposited on a continuous $t_{LN} \simeq \SI{300}{\nano\meter}$ z-cut LN thin film, adhered to a \SI{300}{\nano\meter} optically thick gold back-reflector by a \SI{10}{\nano\meter} chromium adhesive layer. 
The areas covered by semi-transparent gold constitute Fabry-Perot resonators, whose resonances determine the operation wavelength of the device. 
The two-dimensional (2D) Fresnel lenses allow polarization-independent focusing, due to their radial symmetry. For simplicity the polarization is set to be along the x-direction, denoted TM polarization. 
In the design of a Fresnel lens, the relation between wavelength, focal length, and lens dimension is given by $r_m = \sqrt{m\lambda f + \frac{1}{4}m^2 \lambda^2}$, where $\lambda$ is the wavelength of light to be focused, $f$ is the focal length, $m$ is an integer describing the zones, and $r_m$ is the radius of the $m^{th}$ zone. The focal length is a key parameter in the design of a zone plate, and to realize a tight focal spot, a focal length of \SI{40}{\micro\meter} is selected in combination with a wavelength range of 800-\SI{900}{\nano\meter} and a total of $m_{Tot}=19$ zones. This results in a total zone plate radius of $r_{19} \simeq \SI{26}{\micro\meter}$, and a minimum zone width of $\Delta r_{19} \simeq \SI{0.75}{\micro\meter}$. 

The concentric gold rings and the gold back-reflector can serve as integrated metal electrodes for electro-optic tuning of the Fresnel lens (Figure \ref{fig1:Schematic}). The concentric rings of the Fresnel lens are electrically connected by a \SI{2}{\micro\meter} wide wire (Figure \ref{fig1:Schematic}a). 
Applying a voltage across the gold rings and the bottom back-reflector electrode generates an electric field in the sandwiched LN thin film, which induces a change of the refractive index due to the Pockels effect. This shifts the resonance position of the Fabry-Perot resonators, thus giving rise to electrical tunability of the light reflectivity. With the Fabry-Perot optical mode propagating along the $z$-direction, the optical electric field component effectively influencing the Fabry-Perot resonance is in the $x$-direction, thus the relevant electro-optic Pockels coefficient is $r_{13} = \SI{10.12}{\pico\meter/\volt}$ \cite{Jazbinek2002}. The induced change in refractive index is given by $|\Delta n| \simeq \frac{r}{2} n_0^3 \frac{V}{d}$, where $r$ is the relevant Pockels coefficient, $n_0$ is the refractive index, $V$ is the applied voltage, and $d$ is the distance across which the voltage is applied \cite{Pedrotti}. 
To realize an effective AFL, the modulation and reflective properties of the Fabry-Perot resonators are investigated to determine the optimal thickness of the top concentric gold rings, leading to the choice of using a thickness of $t_g=\SI{15}{\nano\meter}$ (see Supporting Information, Section 1). 


\begin{figure*}[!tb]
	\centering
	\includegraphics{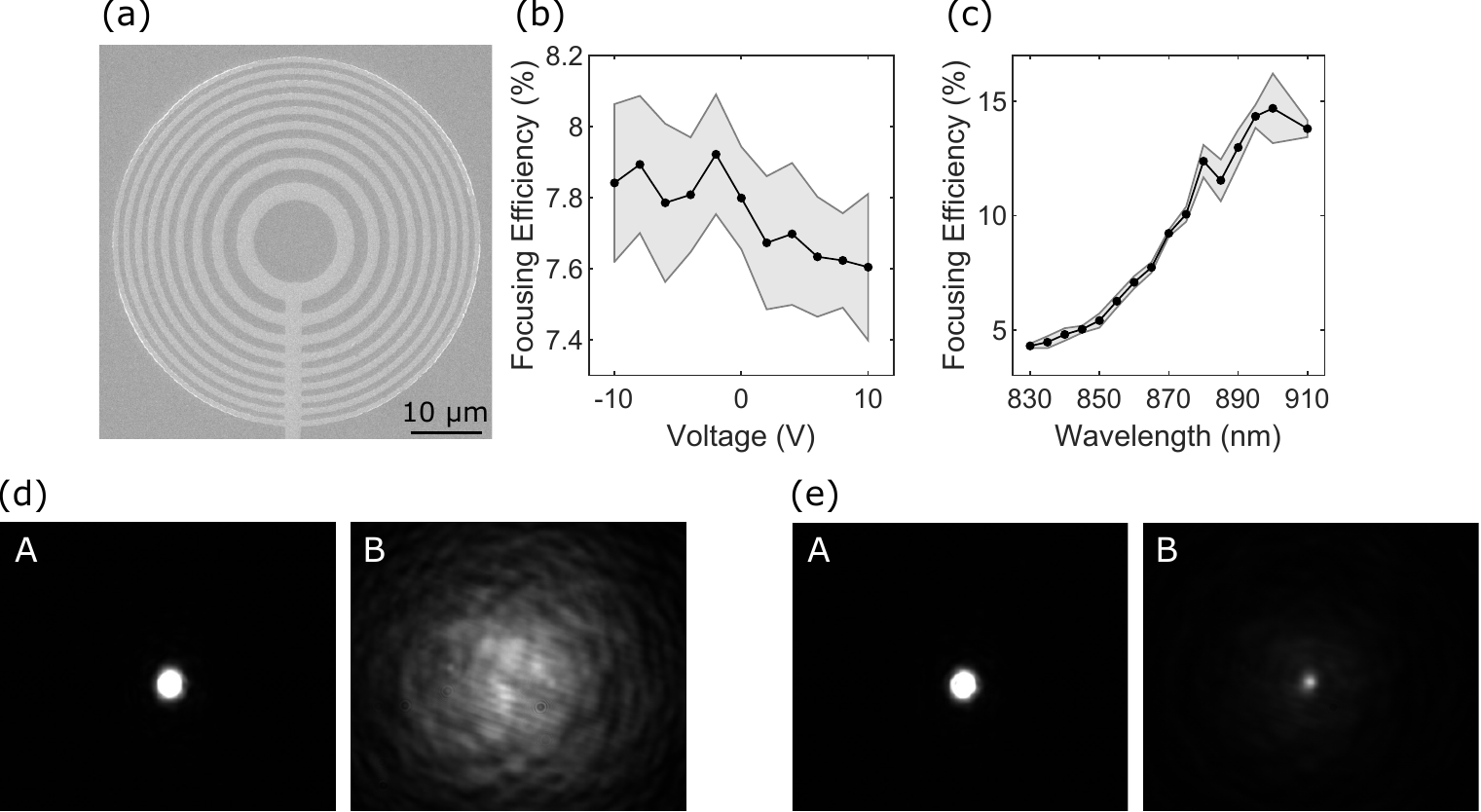}
	\caption{Experimental characterization of the focusing effect of the AFL. (a) Scanning electron microscopy image of the fabricated AFL. (b,c) Verification of the (b) electrical and (c) spectral tunability of the focal spot intensity as a function of DC modulation voltage for a wavelength of \SI{865}{\nano\meter} and of wavelength for a DC modulation voltage of \SI{-10}{\volt}, respectively. The shaded error region represents the linearly interpolated standard deviation of the mean deduced from repeated measurements. (d,e) Optical images in planes A and B (Supporting Information, Figure S4) when the incident light of wavelength \SI{865}{\nano\meter} illuminate (d) flat unstructured gold and (e) the fabricated AFL, respectively. }
	\label{fig3:2DLens}
\end{figure*}


An important characteristic of a focusing element is the focusing efficiency, describing the amount of incident light that is directed to the designed focal spot. Another equally important characteristic when discussing active optical components is the modulation efficiency (calculated as $1-(|I_{min}(\lambda)|/|I_{max}(\lambda)|)$, where $|I_{min}(\lambda)|$ and $|I_{max}(\lambda)|$ are the minimum and maximum achievable intensity at the focal spot for a given wavelength, respectively \cite{Yao2014}), namely the ability to modulate the device performance by applying an external voltage. In this work, we optimize the design to achieve the highest possible modulation efficiency. 
Given the previously mentioned design parameters, the only parameter left to optimize is the design wavelength, which for optimal modulation is determined by calculating the focusing and modulation efficiency as a function of design wavelength, meaning that the zone plate design is adjusted at each iteration of the wavelength (Supporting Information, Figure S3). Simulations show that the focusing efficiency increases significantly from $\sim\! 3$ to $\sim\! \SI{20}{\percent}$ in the investigated wavelength range (see Supporting Information, Section 2). The design wavelength is chosen to be at the point of maximum modulation efficiency, thus $\lambda_0=\SI{815}{\nano\meter}$. Similar simulations are performed for varying incident wavelength but with a constant lens design (Figure \ref{fig2:Simulations}). The performance is equivalent in the vicinity of the design wavelength, and the most significant distinction is for focusing efficiency at longer wavelengths, where the performance loss due to mismatch between the incident and design wavelengths overcomes the otherwise increasing focusing efficiency. The largest differences in focusing efficiencies are observed at the part of steepest slope. The shape of the curve for modulation efficiency resembles those for difference in focusing efficiencies, only slightly blue-shifted, because the modulation efficiency is calculated based on the difference relative to the unmodulated signal. Figure \ref{fig2:Simulations}c shows a scattered field simulation of the investigated AFL at the design wavelength illustrating the focusing ability.


\begin{figure*}[!tb]
	\centering
	\includegraphics{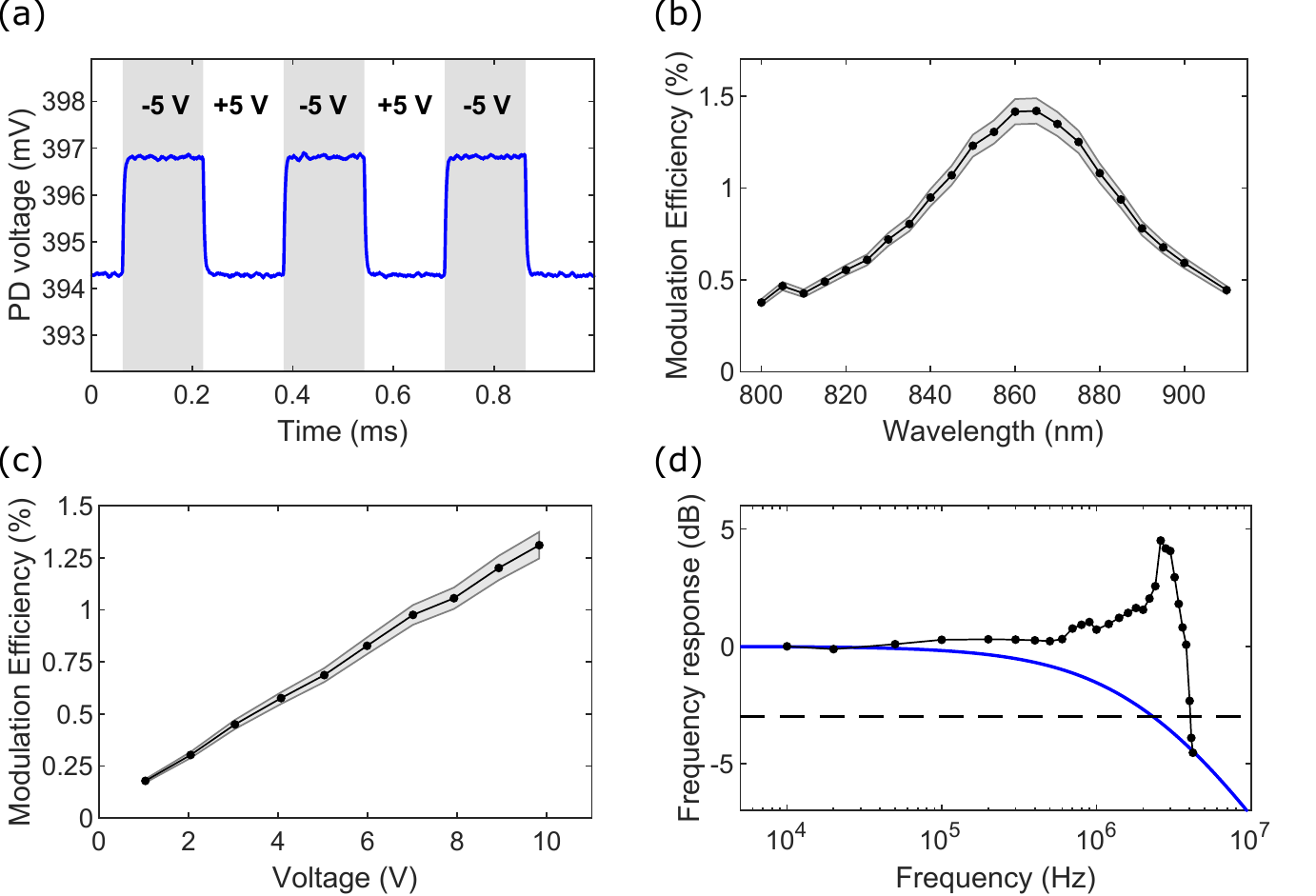}
	\caption{Experimental characterization of the modulation performance of the AFL. (a) Intensity in the focal spot, measured with the photodetector (PD), as a function of time for a wavelength of \SI{865}{\nano\meter}, while the modulation voltage is cycled between \SI{-5}{\volt} and \SI{5}{\volt}, indicated by grey and white backgrounds, respectively. (b) Measured modulation efficiency of the intensity in the focal spot as a function of wavelength for a modulation voltage of \SI{\pm10}{\volt} at a frequency of \SI{3}{\kilo\hertz}. The shaded error region represents the linearly interpolated estimated standard deviation of the mean. (c) Measured modulation efficiency as a function of modulation voltage for a wavelength of \SI{865}{\nano\meter} at a frequency of \SI{3}{\kilo\hertz}. Indicated voltages represent amplitudes of the applied signal. The shaded error region represents the linearly interpolated estimated standard deviation of the mean. (d) Measured frequency response as a function of applied RF signal frequency, normalized to the lowest applied frequency, at a wavelength of \SI{865}{\nano\meter}. The dashed line marks \SI{-3}{\decibel}, and the blue line represents the response of a first order low pass filter with a cutoff frequency of \SI{2.3}{\mega\hertz}, which is calculated as the cutoff of the macroscopic electrodes. Error bars are in the order of data point sizes. }
	\label{fig4:2DModulation}
\end{figure*}


After thorough numerical investigations of the AFL performance, we move on to experimental characterization. An AFL with the chosen design parameters was fabricated using the standard technological procedure based on electron-beam lithography (see Methods). A scanning electron microscopy image of the AFL is shown in Figure \ref{fig3:2DLens}a, showing regular concentric circles without significant fabrication defects. Due to the expected short focal length of the AFL and the relatively low focusing efficiency, it proved difficult to characterize the focusing effect using a conventional imaging setup with near-parallel illumination \cite{Ding2019_PL}, because the focal point will be difficult to distinguish from the interference pattern between incident and normally reflected light from the sample. However, it is possible to verify the focusing effect and determine the focal length by shifting the sample away from the objective from the plane which resulted in a tight focal spot under illumination of flat unstructured gold (plane A in Supporting Information Figure S4) into the plane where the reflected light from the AFL is tightly focused (plane B in Supporting Information Figure S4) \cite{Pors2013,Boroviks2017_PL}. It is deducible from geometrical optics that the distance between these planes is equivalent to twice the focal length. This approach of positioning the sample in planes A and B for radiation incident on flat unstructured gold and the fabricated AFL produces optical images as shown in Figure \ref{fig3:2DLens}d,e, respectively, clearly demonstrating the focusing ability. 

Experimental characterization of the Fabry-Perot modulator shows a measured thickness of the LN thin film of \SI{323}{\nano\meter} (see Supporting Information, Section 1). This corresponds to a deviation of $\sim\!\SI{7.5}{\percent}$ of the nominal thickness, which results in a shift in resonant wavelength of approximately \SI{50}{\nano\meter}. For this reason we see a new wavelength for highest modulation of \SI{865}{\nano\meter} (Figure \ref{fig4:2DModulation}b), which is used as the central wavelength for experimental characterization. This is expected to result in a decrease in performance, as the lens is designed for a wavelength of \SI{815}{\nano\meter}. The measured focal length is $f=\SI{40(2)}{\micro\meter}$. 
Focusing efficiency is investigated as a function of modulation voltage and wavelength (Figure \ref{fig3:2DLens}b,c). The measured and calculated values are not directly comparable as the simulations are for a 2D model. However, the simulations provide trends in the performance for varying voltage and wavelength, which are comparable to the experiments. As is shown by simulations (Figure \ref{fig2:Simulations}a), applying a negative (positive) bias results in an increase (decrease) in focusing efficiency, which is verified by experiments (Figure \ref{fig3:2DLens}b). Similarly, the evolution of focusing efficiency with wavelength (Figure \ref{fig3:2DLens}c) follows that shown by simulations (Figure \ref{fig2:Simulations}a). 



So far, we have characterized the focusing abilities of the AFL, and now we move on to characterize the modulation properties of the intensity in the focal spot (see Methods). The ability to modulate the focal point intensity is visualized by applying an electrical square signal alternating between \SI{\pm5}{\volt}. Measured response of the AFL at an electrical frequency of \SI{3}{\kilo\hertz} shows the dynamic modulation of focusing versus time, and demonstrates as previously stated that a negative bias leads to an increase in focusing efficiency (Figure \ref{fig4:2DModulation}a). Modulation efficiency is measured at a driving voltage of \SI{\pm10}{\volt} for the wavelength range of 800-\SI{910}{\nano\meter} (Figure \ref{fig4:2DModulation}b). The maximum modulation efficiency of \SI{1.5}{\percent} is measured at a wavelength of \SI{865}{\nano\meter}, and the measured dispersion of the modulation efficiency is in agreement with the simulated wavelength dependence (Figure \ref{fig2:Simulations}b). A linear relation is expected between modulation efficiency and voltage due to the previously stated formula for induced refractive index change, and the resulting shift in the wavelength of Fabry-Perot resonance. This relation is verified by experimental characterization (Figure \ref{fig4:2DModulation}c). 
The electro-optic frequency response is characterized from \SI{10}{\kilo\hertz} to \SI{4.5}{\mega\hertz} (Figure \ref{fig4:2DModulation}d). The device frequency response exhibits an increase in performance for larger signal frequency before abruptly dropping, resulting in a \SI{-3}{\decibel} cutoff frequency of \SI{4}{\mega\hertz}. Frequency response fluctuations might be attributed to piezoelectric resonances in LN, and the accompanied variations of the permittivity and the electro-optic activity in LN when the crystal strain becomes unable to follow the external electric field (clamped crystal response) \cite{Takeda2012,Jazbinek2002,Thomaschewski2020}. Using a simple parallel plate capacitor formula, the capacitance of the device and electrodes is calculated to be $C_c = \SI{1.36}{\nano\farad}$, which corresponds well with the measured capacitance of $C_m \simeq \SI{1.5}{\nano\farad}$. Assuming a \SI{50}{\ohm} resistive load ($f=1/[2\pi RC]$), the calculated \SI{-3}{\decibel} cutoff frequency is \SI{2.3}{\mega\hertz}, which is indicated by a first order low pass filter response (blue line of Figure \ref{fig4:2DModulation}d), intersecting the measured data just below the \SI{-3}{\decibel}-line. Disregarding the macroscopic electrodes and electrical wiring, the capacitance of the device is calculated to be \SI{0.83}{\pico\farad}, resulting in a cutoff frequency of \SI{3.8}{\giga\hertz}, which is easily supported by the fast electro-optic Pockels effect. Thus the electrical bandwidth can be considerably improved by optimizing the macroscopic electrodes and electrical wiring. 


\section{Conclusion}
In summary, we have presented and experimentally investigated an approach to realize a flat electrically tunable Fresnel lens by utilizing the electro-optic effect in a thin lithium niobate layer sandwiched between a continuous thick bottom and nanostructured top gold film serving as electrodes. We have designed, fabricated and characterized the active Fresnel lens that exhibits focusing of 800-\SI{900}{\nano\meter} radiation at the distance of \SI{40}{\micro\meter} with the focusing efficiency of \SI{15}{\percent} and modulation depth of \SI{1.5}{\percent} for the driving voltage of \SI{\pm10}{\volt} within the bandwidth of \SI{4}{\mega\hertz}. It should be noted that the modulation efficiency can significantly be improved by using a high-quality top gold film with the optimal thickness of \SI{12}{\nano\meter} (see Supporting Information, Section 1), as the currently used \SI{15}{\nano\meter}-thin gold film is likely to be inhomogeneous (island-like). Furthermore, redesigning the macroscopic electrodes and electrical wiring can considerably improve the electrical bandwidth reaching the GHz range as discussed above. In comparison with other electrically tunable thin lenses \cite{Park2020,Shirmanesh2020,vandeGroep2020}, the configuration presented here is attractive due to its simplicity in design and fabrication and inherently fast electro-optic response (see Supporting Information, Section 4). Overall, we believe the introduced electro-optic metasurface concept is useful for designing dynamic, electrically tunable flat optics components. 


\section{Methods}
\textit{Modeling.} 
Simulations are performed in the commercially available finite element software \textit{COMSOL Multiphysics}, ver. 5.5. Fabry-Perot modulators and Fresnel lenses are modulated to determine reflectivity and focusing properties. All simulations are performed for 2D models, due to computational restraints. In all setups, the incident wave is a plane wave traveling downward, normal to the sample. Interpolated experimental values are used for the permittivity of gold \cite{Johnson1972}, LN \cite{Zelmon1997}, and chromium \cite{Johnson1974}, and the medium above the sample is air. 
For simulation of the Fabry-Perot modulators, periodic boundary conditions are applied on both sides of the cell, while the top and bottom boundaries are truncated by ports to minimize reflections. The top port, positioned a distance of one wavelength from the top electrode, handles wave excitation and measures complex reflection coefficient. For simulation of the AFL, periodic boundary conditions are applied on one side, so it is only necessary to model half the zone plate. All other boundaries are truncated by scattering boundary conditions, also to eliminate reflections. Focusing efficiency is determined by integrating the reflected power over an area corresponding to twice the beam waist of a Gaussian beam focused at the focal point and dividing by the incident optical power. 

\textit{Fabrication.} 
Fabrication of the AFL is done using a combination of nanostenciling and electron beam lithography and lift-off. A substrate with the following layered structure is obtained commercially: Bulk LN substrate, \SI{3}{\micro\meter} SiO\textsubscript{2}, \SI{30}{\nano\meter} of chromium, \SI{300}{\nano\meter} of gold, \SI{10}{\nano\meter} of chromium and lastly a \SI{300}{\nano\meter} thin film of LN (NANOLN). Initially, macroscopic electrodes are deposited by thermal evaporation of \SI{3}{\nano\meter} titanium and \SI{50}{\nano\meter} gold through a shadow mask. Subsequently, $\sim \! \SI{200}{\nano\meter}$ of PMMA 950K A4 is spin-coated, and the Fresnel zones and modulator squares are exposed at \SI{30}{\kilo\volt} using electron beam lithography. Alignment between the macroscopic electrodes and optical devices is performed manually. After development, the devices are formulated by thermal evaporation of \SI{1}{\nano\meter} titanium and \SI{15}{\nano\meter} gold followed by lift-off in acetone. The fabricated modulator squares are \SI{100x100}{\micro\meter}, and the AFL has a radius of \SI{26.1}{\micro\meter} and consists of 19 zones, with even zones formed by gold deposition. 

\textit{Electro-optical characterization.} 
During fabrication, the concentric rings are interfaced to macroscopic electrodes. For electro-optical characterization, the sample is mounted on a home-made sample holder, that connects to the macroscopic top electrode, and electrical connection to the bottom electrode is obtained by applying a conductive paste on the edge of the sample. 
The incident light is a low power, continuous-wave laser beam from a tunable laser, which is focused by a 50X objective to form a tightly focused spot in plane A (Supporting Information, Figure S4) on flat unstructured gold. The reflected light is collected by the same objective, separated from the incident light by a beam splitter and viewed on a camera. Focusing efficiency is determined as the ratio of focused light from the device viewed in plane B to the amount of reflected light on flat unstructured gold viewed in plane A. 
For characterization of the modulation properties, the focal spot is manually isolated with an iris and the camera is replaced with a photodetector connected to an oscilloscope. RF modulation signals are supplied by a function generator, and modulation of the focal spot intensity is observed on the oscilloscope.

\begin{acknowledgement}

The authors thank for financial support from Villum Fonden (Award in Technical and Natural Sciences 2019 and Grant No. 00022988 and 37372). \\
C.D.-C. acknowledges advice from Chao Meng on the optical characterization of focusing devices.

\end{acknowledgement}

\begin{suppinfo}

The following files are available free of charge \\ 
\\
\textbf{Supporting Information.} Investigation of the Fabry-Perot modulator, determination of the design wavelength, setup for electro-optical characterization, and comparison of electrically tunable thin lenses 

\end{suppinfo}

\section{Author contributions}
S.I.B. conceived the idea. C.D.-C. designed the sample and performed the numerical simulations with F.D.. C.D.-C. and M.T. fabricated the structures and conducted the electro-optical characterization. C.D.-C. analyzed the results, which were discussed by all authors. C.D.-C. and S.I.B. wrote the manuscript with revisions by all authors. S.I.B. supervised the project. 

\bibliography{ActiveFZP_BIB.bib}


\end{document}